\documentclass{article}
\usepackage{spconf,amsmath,graphicx}
\usepackage{hyperref}
\usepackage{float} 
\usepackage{pifont}
\newcommand{\cmark}{\ding{51}}%
\newcommand{\xmark}{\ding{55}}
\usepackage{svg}
\usepackage{tablefootnote}
\usepackage{bibspacing}
\ninept

\title{A dataset for Audio-Visual Sound Event Detection in Movies}

\name{Rajat Hebbar, Digbalay Bose, Krishna Somandepalli, Veena Vijai, Shrikanth Narayanan}
\address{Signal Analysis and Interpretation Laboratory, University of Southern California}
\begin{document}
%
\maketitle

\begin{abstract}
Audio event detection is a widely studied audio processing task, with applications ranging from self-driving cars to healthcare.
In-the-wild datasets such as Audioset have propelled research in this field. However, many efforts typically involve manual annotation and verification, which is expensive to perform at scale.
Movies depict various real-life and fictional scenarios which makes them a rich resource for mining a wide-range of audio events. 
In this work, we present a dataset of audio events called Subtitle-Aligned Movie Sounds (SAM-S).
We use publicly-available closed-caption transcripts to automatically mine over 110K audio events from 430 movies.
We identify three dimensions to categorize audio events: \textit{sound}, \textit{source}, \textit{quality}, and present the steps involved to produce a final taxonomy of 245 sounds. 
We discuss the choices involved in generating the taxonomy, and also highlight the human-centered nature of sounds in our dataset.
We establish a baseline performance for audio-only sound classification of 34.76\% mean average precision, and show that incorporating visual information can further improve the performance by about 5\%.
Data and code are made available for research at \href{https://github.com/usc-sail/mica-movie-audio-events}{https://github.com/usc-sail/mica-movie-audio-events}
\end{abstract}
\begin{keywords}
Audio Event Detection, Movies, Audio Recognition, Audio Visual Dataset
\end{keywords}

\section{Introduction}
\label{sec:intro}
Audio events are naturally occurring non-verbal sounds produced by humans/objects.
Robust detection of such audio events can reveal information about one's acoustic environment, their psychological state, and help automate rich transcription of multimedia data.
Audio event detection (AED) is used in a wide range of domains, including context-aware smart device applications such as in smart-phones \cite{appleSound}, smart-speakers \cite{amazonSound} and self-driving cars \cite{nandwana2016towards,mesaros2017dcase}, acoustic monitoring for health and well-being applications \cite{jati2021temporal,goetze2012acoustic} as well as large-scale multimedia indexing \cite{gemmeke2017audio,ma2021computational}.
Recently, the introduction of large-scale ``in-the-wild" datasets such as Audioset \cite{gemmeke2017audio} and VGGSound \cite{chen2020vggsound} has enabled prolific AED research. 
Neural-network representations learned over Audioset have been used for several audio classification tasks such as emotion recognition \cite{kong2020panns}, gender identification \cite{hebbar2018improving} and music classification \cite{kong2020panns,Ramrez2019MachineLF}. However, curating such datasets usually involves manual intervention at multiple stages -- during data collection, labeling and taxonomy generation. 
Furthermore, data collected from YouTube sources are subject to attrition due to videos being taken down or made private. 

\begin{figure}[t]
  \centering
  \includegraphics[width=\linewidth]{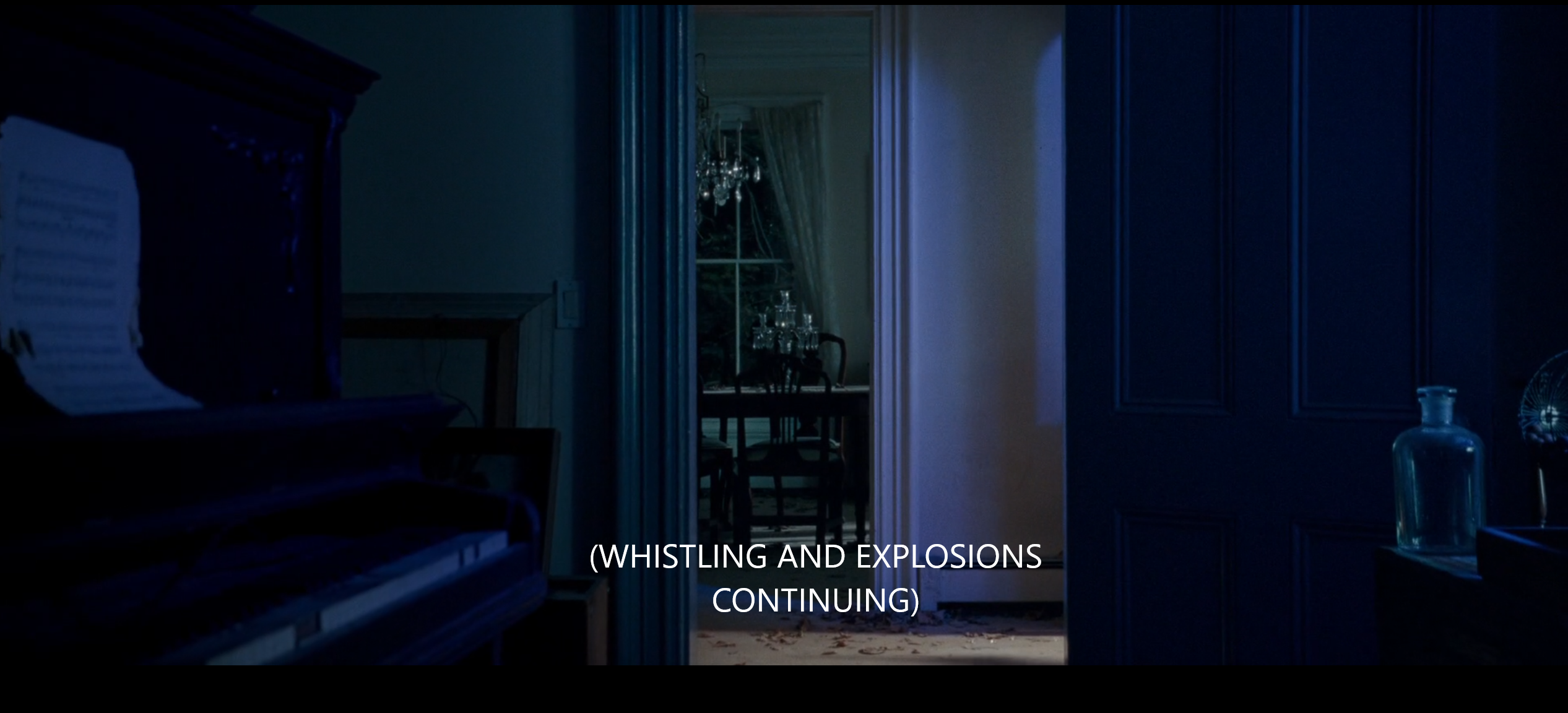}
   
  \caption{Closed caption showing audio events occuring off-screen in a movie}
  \label{fig:sub}
  \vspace{-3mm}
\end{figure}

As \textit{sound effects}, audio events form an integral component of the movie audio stream. 
Deliberate placement of sounds and background-score in a movie scene helps construct a rich narrative and elicit the intended emotional response from viewers.
While a large fraction of sound-effects in movies are naturally produced (human/animal vocalizations, music, etc.), some sounds, known as ``Foley sounds" \cite{nozaic2006introduction} are added in post-production.
Foley involves the use of `everyday' objects to create sound effects that imitate naturally occurring audio events in different ambient environments; be it the use of snapping \textit{celery} to imitate the sound of breaking bones, or popping the bottom of \textit{trashcans} to amplify the sound of heartbeats\footnote{\href{https://blog.storyblocks.com/inspiration/foley-sfx-everyday-household-objects/}{blog.storyblocks.com/inspiration/foley-sfx-everyday-household-objects}}. Foley is an effective tool that enables simple and inexpensive reproduction of such sounds. It allows for the possibility of audio events that may not be commonly found in the aforementioned data sources (e.g., vehicles crashing, light footsteps, gunshots). 
Furthermore, it was shown that Foley sounds are nearly indistinguishable from their naturally produced counterparts \cite{de2013real}, which makes it useful for developing AED models.

Closed-captions (CC) are time-aligned transcriptions of character dialogues and sound effects.
These captions are mandated for several broadcast media, including movies and TV-shows, in an effort to make media more accessible to the hearing-impaired and non-native speakers.
Following the guidelines provided by the Described and Captioned Media Program (DCMP) \cite{Captioni46:online}, audio captioners are expected to label audio events that are deemed relevant to the plot of the movie/TV-show.
Therefore, existing captions can be used to obtain audio-events from movie data in a precise manner.

The contributions of our work are three-fold: \\
1. We use simple and scalable methods to automatically extract audio captions from movies and categorize an audio event along the dimensions of \textit{sound}, \textit{source} and \textit{quality}.\\
2. We propose a flat taxonomy of sounds, decoupled from their sources. 
Unlike previous AED taxonomies, this enables us to group together acoustically similar sounds from different sources.\\
3. We leverage visual-cues using early multimodal-fusion of audio and video features in a transformer setup to establish baseline audio event detection performance on our dataset.

The rest of the paper is organized as follows:
Section \ref{sec:related} discusses existing resources and methodology.
In Sec \ref{sec:dataset}, we outline the taxonomy generation process using subtitle tags and compare with audioset taxonomy
In Sec.~\ref{sec:baseline}, we describe the methods used to develop baseline AED models on our dataset.

\section{Related work}
\label{sec:related}
\textbf{Data Resources:}
Audioset \cite{gemmeke2017audio}, one of the first large-scale AED datasets, includes over 2-million clips with weakly-tagged audio-event classes.
The audioset ontology, is the most comprehensive taxonomy of audio-events, comprising 527 different audio-events in a hierarchical structure based on the source of an audio-event.
Human-raters were used to label the Audioset data at clip-level.

In order to reduce the manual effort involved in labeling, VGGSound \cite{chen2020vggsound} dataset was proposed, which used a scalable pipeline of mining \textit{visually-grounded} audio events from YouTube. 
Existing machine learning models were used to automatically verify presence of visual signature, and to reject possible false-positive audio classes during data curation.
However, a shortcoming is that such methods still do not guarantee occurrence of a tagged sound-event, allowing for some label-noise in exchange for reduced manual effort.
Furthermore, it is often the case that an audio-event is heard but not shown on screen, a scenario that is not covered by the VGGSound dataset.

FSD50K \cite{fonseca2022fsd50k} consists of over 50K audio events collected in-the-wild, which are annotated across 200 audioset classes on the freesound platform \cite{fonseca2017freesound}.
Apart from these, several smaller-scale AED datasets exist such as Mivia \cite{foggia2015reliable}, DESED \cite{turpault2019sound}, UrbanSound8k \cite{salamon2014dataset}. 
These are typically targeted toward a specific subset of sounds such as indoor, outdoor and rare audio events.

\begin{table}[ht]
\centering
\caption{Details of different audio event detection datasets. Here, SL refers to whether the audio event labels are precise or are weakly-labeled, PA refers to whether the dataset is publicly available or not; FSD - Freesound platform.}
\label{table:datasets}
\resizebox{\columnwidth}{!}{
\begin{tabular}{|l|l|l|l|l|l|l|}
\hline
\textbf{Dataset}  & \textbf{Domain}  & \textbf{Clips} & \textbf{Classes} & \textbf{SL}         & \textbf{PA}  & \textbf{Annotation} \\ \hline
Mivia \cite{foggia2015reliable} & Synthetic & 6K      & 3         & \cmark & \cmark & Manual\\ \hline
UrbanSound8k \cite{salamon2014dataset} & FSD & 8.7K & 10 & \cmark & \cmark & Manual \\ \hline
DESED \cite{turpault2019sound} & FSD & 12K   & 10 &  \cmark &  \cmark & Manual \\ \hline
FSD50K \cite{fonseca2022fsd50k} & FSD & 50K & 200 & \xmark & \cmark & Manual \\ \hline
Audioset \cite{gemmeke2017audio} & Youtube & 2.1M      & 527         & \xmark & \xmark\tablefootnote{Youtube policy: \href{https://www.youtube.com/static?gl=US\&template=terms}{https://www.youtube.com/static?gl=US\&template=terms}} & Manual\\ \hline
VGGSound \cite{chen2020vggsound} & Youtube & 200K      & 309         & \xmark & \xmark  & Automatic\\ \hline
SAM-S    & Movies  & 110K      & 191         & \cmark & \cmark & Semi-automatic\\ \hline
\end{tabular}
}
\end{table}

\noindent \textbf{AED:}
Until recently, convolutional (CNN) models have been  used widely for AED. 
A light-weight version of VGG-16, called \textit{VGGish} \cite{hershey2017cnn} was the first benchmark model for AED on Audioset. 
Several commonly used CNN architectures were compared on Audioset \cite{kong2020panns}, and it was shown that light-weight CNNs can obtain comparable performance to their larger counterparts.
More recently, transformer architectures such as VATT \cite{akbari2021vatt} and AST \cite{gong2021ast} have shown state-of-the-art (SoA) performance for
audio-only AED. 
VATT used a self-supervised contrastive loss to pretrain general multimodal representations, which were then finetuned for AED.
AST incorporated state-of-the-art vision transformers \cite{touvron2021training} in AED using spectrogram features.

There have been a few audio-visual methods proposed for audio event detection and localization.
Mid-level attention based fusion was used on audio-visual streams with CNN backbones \cite{fayek2020large}.
An optimal multimodal fusion mechanism, called \textit{gradient-blending} \cite{wang2020makes}, was proposed to address variable overfitting rates across modalities.
More recently, attention-bottlenecks in multimodal transformer architectures have been proposed \cite{nagrani2021attention}, showing SoA AV-performance on Audioset using early-fusion.
Cross-modal attention mechanisms have been used for audio-visual localization of event sources, and weakly-supervised detection \cite{brousmiche2020intra,wu2019dual,xuan2020cross}. 

In the context of multimedia, there have been limited works that have analysed audio events.
Gunshot and explosion classification was studied based on dictionary learning from MFCC features \cite{Penet2013AudioED}.
Audio event change detection was explored via clustering methods in a set of 8 movies \cite{yang2013audio}.

\begin{figure}[t]
  \centering
  \includegraphics[width=\linewidth]{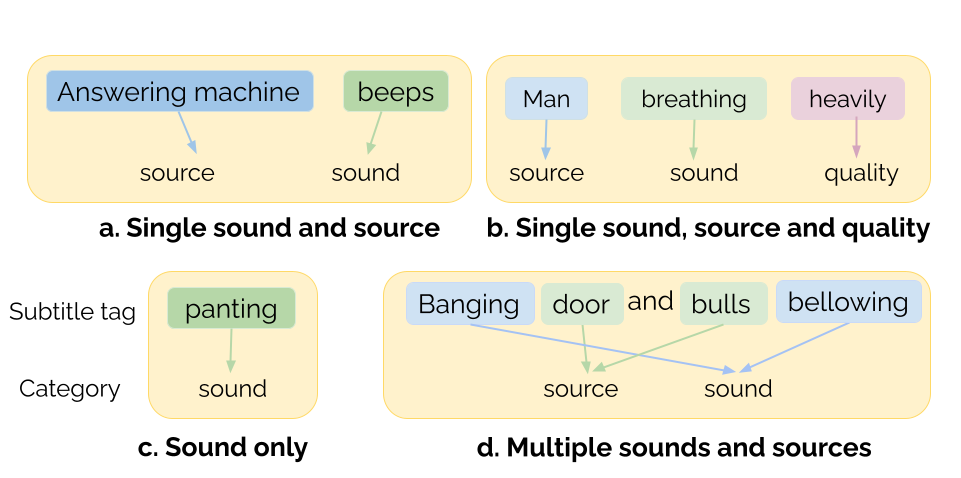}
  \caption{Annotation examples for sound, source and quality categorization in movie audio events}
  \label{fig:sub_annot}
  \vspace{-3mm}
\end{figure}

\section{Subtitle Aligned Movie Sounds (SAM-S)} 
\label{sec:dataset}
The SAM-S dataset we introduce in this work comprises 430 top-grossing Hollywood films from the years 2014 to 2018.
In order to mine audio events, subtitle files for each of the movies were obtained automatically\footnote{\href{https://github.com/ruediger/VobSub2SRT}{https://github.com/ruediger/VobSub2SRT}}. 
Closed-captioned subtitles extracted in this manner contain time-aligned character dialogues and plot-relevant event tags. 
It is important to note here that closed-captions are not exhaustive in labeling all audio events that occur in a movie, i.e., the tagging process has low-recall.
However, the tagged captions are accurate in terms of the labeled sound, i.e., high-precision. 
This precision-recall trade-off means that while we lose potentially useful data, we ensure minimal additional human effort for annotation and cleanup, and a large-enough set of sounds to develop and evaluate AED models.

These tags are typically enclosed in braces. We automatically extract these tags and the associated time-stamp from the subtitle file. 
In total, we obtain just over 116K subtitle tags, of which 20,817 are unique. 
These subtitle tags are descriptive in nature, and include information about the sound, source and quality of the audio event occurring (see Figure \ref{fig:sub_annot}).
While information about sound is always present, often the source and quality of the audio event are not tagged.
In fact, following DCMP guidelines, captioners are expected to label the source of the sound with the exception of the instances where the source is clearly visible on-screen. 
Out of the 21K unique tags, 1.5K are unigrams - usually indicating only the sound, 11K are bigrams - which include the sound and source, and the rest are n-grams, $n\ge3$, which could refer to the quality of the audio event or multiple simultaneously occurring events.
Due to the presence of source-ambiguous audio events in our data, we chose to adopt a flat taxonomy as opposed to a hierarchical one as in \cite{gemmeke2017audio}, which we discuss in more detail in Sec.~\ref{sec:structure}.

\subsection{Taxonomy generation}
\label{sec:taxonomy}

The following steps outline the procedure to label, refine and condense our final taxonomy:
\noindent
\textbf{Categorization:} We conducted an annotation task using Mechanical Turk in order to categorize a given subtitle tag into `source', `sound' and `quality' classes. 
A few sample examples of the annotation task is provided in Figure \ref{fig:sub_annot}. 
Annotators were explicitly asked to only sample from the words in the tag, and not introduce/interpolate from context (f.e, a subtitle tag of "flickering", would only have a sound of "flickering" and no source). 
Three annotations were used for each tag, and majority voting was used for each category. 
All ties were resolved by an author. 
For this task, we chose the set of subtitle tags that occur at least 5-times in the dataset. 
This set of 2161 tags covers 80\% ($\sim$91K) of the audio-visual events that occur in the dataset. 

\begin{figure}[t]
  \centering
  \includegraphics[width=\linewidth]{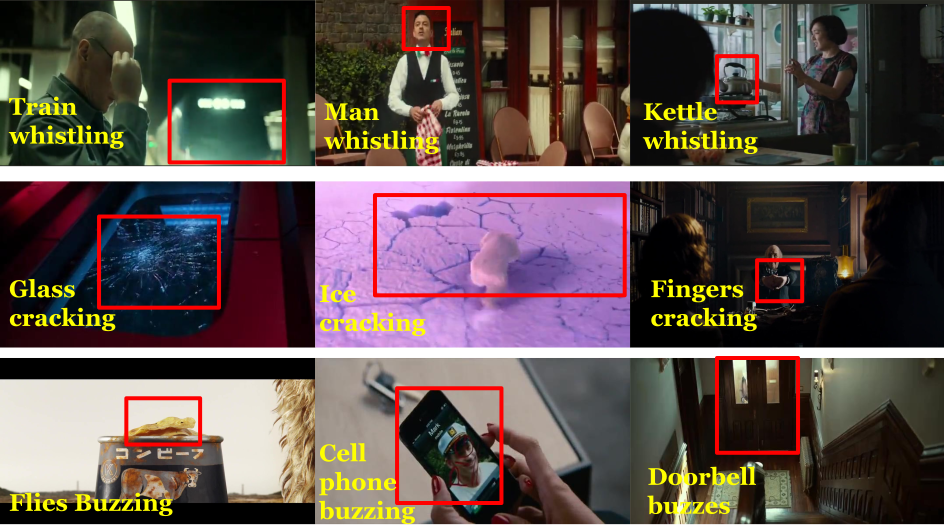}
  \caption{Examples of sounds originating from multiple sources in movies: (Top) "Whistling" sound from different sources in SAM-S, (Middle) "Cracking", (Bottom) "Buzzing" }
  \label{fig:sounds_sources}
  \vspace{-3mm}
\end{figure}

\noindent
\textbf{Lemmatization:} For each of the categories, the set of annotations obtained were lemmatized using an opensource NLP-toolkit - spaCy\footnote{\href{https://spacy.io/api/lemmatizer}{https://spacy.io/api/lemmatizer}}. 
The lemmatization process was manually verified and errors were corrected by an author.
Following this, a total of 254 sources, 254 sounds and 115 qualities form the initial taxonomy of our dataset.
\noindent
\textbf{Automatic tagging:} 
Next, we created a dictionary mapping the original words in the subtitle tags to the transformed version for each of the categories. 
Using this dictionary, we attempt to automatically label the $\sim$25K tags which were left out of the manual annotation process due to low frequency.
In cases where both sound and source were detected, an additional check was added to ensure that the sound-source combination was seen in the manual annotation-scheme. 
New combinations were disregarded as labeling error and such samples were not used. 
Any audio event without a sound tagged was discarded.
We were able to automatically tag around 10K more sounds and 5K sources in this manner.

\textbf{Label set refinement:}
We do a final manual pass of the unique sound and source tags and combine classes that were not taken care of by the lemmatization, e.g., "laugh" and "laughter", "explode" and "explosion", and "thunder" and "thunderclap".
We use a named entity recognizer\footnote{\href{https://spacy.io/usage/linguistic-features\#named-entities}{https://spacy.io/usage/linguistic-features\#named-entities}} to detect names of persons and merge into a single source class.
The resultant dataset consists of 95,452 samples covering 101,311 sounds from 245 classes, 21,460 sources from 183 classes and 7212 qualities from 93 classes.


\begin{figure}[t]
   \makebox[\linewidth][c]{\includegraphics[width=\linewidth]{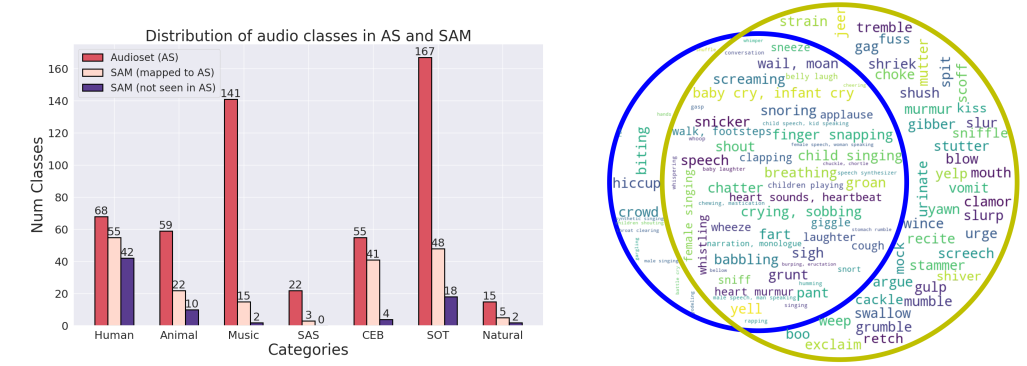}}
  \caption{a) Number of classes in each category in the Audioset taxonomy, b) Venn diagram for human-centered audio events in Audioset (Blue) and SAM-S (Yellow)}
  \label{fig:taxonomy}
  \vspace{-3mm}
\end{figure}

\subsection{Acoustic and Semantic grouping of sounds} 
\label{sec:structure}
Several audio event "sounds" in our dataset can be associated with multiple distinct sources (See Fig. \ref{fig:sounds_sources}). 
For example, the sound of buzzing is associated with different sources - flies/insects, cell-phone, doorbell and alarm, in our dataset.
The sounds of a cell-phone buzzing and an alarm buzzing can be considered to be acoustically similar, so can the sounds of flies and bees buzzing. 
However, the buzz sound of a fly/insect has distinct acoustic signature related to its frequency spectrum and timing characteristics that distinguishes it from the cell-phone or alarm buzzes. 
In a semantic sense, these are all generally referred to as buzzing.
Hence, for modeling purposes, one of two options can be considered: 
1) Retain source-specific sounds as individual classes.
2) Merge acoustically and semantically similar sounds 

If such sounds are considered as a single class, we reduce the total number of classes and obtain more representative samples per class, while at the same time increase the acoustic variability within a single sound class.

As an example, the audioset (AS) taxonomy \cite{gemmeke2017audio} is hierarchical, with the different branches of the hierarchy being organized by the source.
Here, the sound "buzz" is seen in 5 different audioset hierarchies, under `alarm', `telephone', `fly', `bees' and `onomatopoeia'. 
Most modeling techniques developed on the Audioset data adopt option 2, by flattening out the hierarchy and considering each sound as a single class.


In our taxonomy, we make a practical choice of not following the audioset method due to two reasons: 1. Keeping source-ambiguous sounds separate significantly reduces the number of samples available to train/evaluate machine learning models, 2. Following DCMP guidelines, we do not always have information about the source of an audio-event.

\subsection{Overlap with Audioset}
We are interested in understanding the distribution of sound events in movies and how they compare with existing datasets. 
In order to do this, we distribute each of the sound classes in our dataset into two groups, a) shared sounds, b) movie-specific sounds. 
Shared sounds refer to the classes which exist in both our taxonomy as well as the Audioset taxonomy while movie-specific sounds are those that exist in our taxonomy alone.

For each of the sounds in SAM-S, we matched one or more corresponding classes in Audioset taxonomy \cite{gemmeke2017audio}, in order to analyze label coverage.
Pairwise cosine similarity scores were extracted between sentence transformer embeddings (MiniLM-L6-v2) \cite{reimers-2019-sentence-bert} of the two taxonomies.
For each sound, top-5 audioset class matches were then manually verified.
Out of 245 sound classes, we found one or more direct matches in Audioset for 170 classes. 
The remaining classes were manually mapped, if found relevant, to an equivalent AS class. 
Sounds originating from multiple possible sources were mapped to each of the relevant categories.
This resulted in a total of 189 shared sound classes.
It is interesting to note that although some of these classes exist in the audioset taxonomy, they have no representative data samples, for example; screech, blare, yawn and booing.

Finally the remaining `movie-specific sounds' are manually grouped into the high-level categories.
The distribution of the `shared sounds' and 'movie-specific sounds' are shown in Fig.~\ref{fig:taxonomy}a. 
We can see that highest coverage is obtained for human sounds and source-ambiguous sounds. 
In the case of music, instances in SAM-S typically do not specify music instruments/genres, hence we do not observe the level of detail as in Audioset.

In Fig.~\ref{fig:taxonomy}b, we can see that SAM-S includes most of the human sound classes found in Audioset. 
However, it also contains many sounds not found in Audioset; including single-person sounds such as yawn, sniffle, strain and scoff, as well as crowd-sounds such as murmur, clamor and argue. 
Therefore, SAM-S can also be used to augment Audioset with more fine grained human-centric sounds.

\begin{figure}[t]
  \centering
  \includegraphics[width=\linewidth]{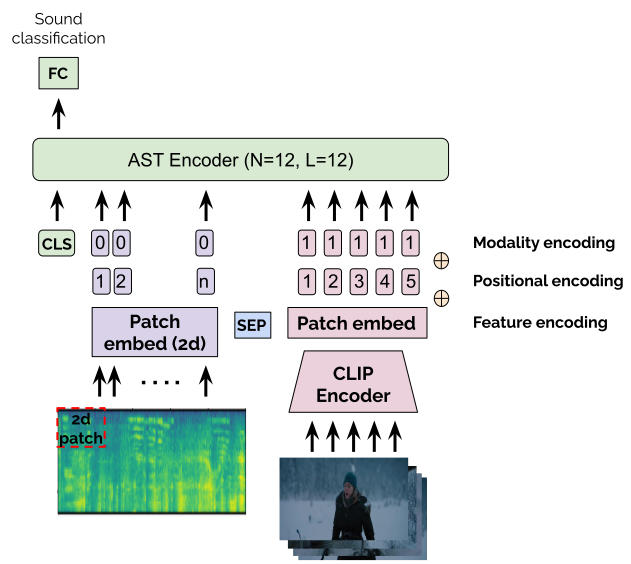}
  \caption{AST-MM: Multimodal Transformer architecture for audio event detection in movies.}
  \label{fig:aed_model}
\end{figure}

\section{Experiments}
\label{sec:baseline}
In order to transfer knowledge from large-scale models, we create 10s segments by adding context ($\sim$5s) on either side of the audio-event. 
We create a train-validation-test (80-10-10) split based on the movies, i.e., we use 344 movies for training, 43 for validation and 43 for final evaluation.
For development and evaluation purposes, we restrict to sound classes that explain at least 0.1\% of the entire data, which results in 120 sound classes.

\subsection{Baseline Audio Models}
We conduct audio-only baseline experiments using two SoA models. 
The first is a Resnet-18 model \cite{chen2020vggsound} using 512-dim log-spectrograms pre-trained on VGGSound dataset. 
We also fine tune a transformer-based AED model - AST \cite{gong2021ast}, which has shown state-of-the-art performance for audio event detection on Audioset.
128-dim log-mel spectrograms are used as features to AST.
For augmentation, we use mixup \cite{tokozume2017learning} with probability 0.5 and sample the mixup-lambda from a Beta-distribution with parameters $\alpha$=$\beta$=10. We also use SpecAugment \cite{park2019specaugment} with a time-frequency mask of 192x48.

For our experiments, we use a batch size of 20, initial learning rate of 1e-5 and a multi-step learning rate scheduler at epoch 5 and 25 with decay of 0.85 similar to as in \cite{gong2021ast}. 
We train each model for 30 epochs.  
Since our classes are multi-label and we use mix-up, we use binary cross-entropy loss.
As evaluation metrics, we use mean average precision (mAP), area under curve (mAUC), and d-prime.

\subsection{AST-MM}

As visual features, we use the output of CLIP-encoder (ViT-B/32) \cite{Radford2021LearningTV}.
CLIP was trained in a contrastive manner using language-image pairs and has been widely used in a number of image recognition tasks.
We also chose CLIP because of its ability to generalize well to unseen objects and scenes, which is often the case in movies.
We extract 512-dim CLIP features at 1fps, and pad/crop the resulting features to a sequence length of 12.

We use position embeddings and modality-specific segment embeddings to encode multiple modalities in a transformer setup as in previous work\cite{kiela2019supervised}.
We pass CLIP features through a linear layer to match-dimensions of the audio patch embeddings (768-dim).
We experiment with three different position embeddings: 1. Fixed sinusoidal position embeddings (S) \cite{vaswani2017attention}, 2. Learnable embeddings initialized with pretrained BERT position embeddings (B) \cite{devlin2018bert}, 3. Learnable, randomly initialized.

We also use a separator token to distinguish audio and visual sequences as we find it helps empirically.
The final input representation to the encoder is obtained by adding the patch, position and modality embeddings for each of the sequences.

\subsection{Results}
From Table \ref{table:results}, we see that the audio-only model trained on Audioset clearly outperforms the one trained on VGGSound. 
Apart from the size of the datasets and model architecture, a reason for this could be that the constraint on audio-visual correspondence limits the range of sound classes seen in VGGSound, hence affecting its transferability to other domains.
The multimodal AST-MM model shows a 5\% relative improvement over the audio-only model.

\begin{table}[t]
\caption{Uni- and multi- modal results on SAM-C}
\label{table:results}
\begin{tabular}{|l|l|l|l|l|} \hline
\textbf{Model} & \textbf{Modality}    & \textbf{mAP} & \textbf{mAUC} & \textbf{d-prime} \\ \hline
VGGSound      & A &   14.1          &   87       &     1.59    \\ \hline
AST              & A  & 34.76   & 95.02         & 2.33    \\ \hline
AST-MM (S) & AV & 35.67  & 95.05   &  2.33 \\ \hline
AST-MM (B) & AV & 35.82     & 95.11   & 2.34 \\ \hline
AST-MM (L) & AV & \textbf{36.3 }    & \textbf{95.25}   & \textbf{2.36} \\ \hline
\end{tabular}
\end{table}

\section{Conclusion}
In this paper, we release a dataset curated for audio-event detection in movies. 
We describe a scalable method for generating a flat-taxonomy for audio events, and compare it with existing taxonomy popularly used for audio event detection.
We designed an annotation scheme to categorize the sound and source of an audio event, solely from subtitle tags. 
We employ state of the art machine learning models to establish baseline AED performance on the SAM-S corpus.
We incorporate visual information using CLIP-encoder features in a early-fusion manner to further improve AED multimodally.

\footnotesize
\bibliographystyle{IEEEbib}
\bibliography{main}

\end{document}